# Privacy-Preserving Healthcare Data in IoT: A Synergistic Approach with Deep Learning and Blockchain

**Behnam Rezaei Bezanjani[1] . Seyyed Hamid Ghafouri[1*] . Reza Gholamrezaei[1]**

**Abstract:** The integration of Internet of Things (IoT) devices in healthcare has revolutionized patient care by enabling real-time monitoring, personalized treatments, and efficient data management. However, this technological advancement introduces significant security risks, particularly concerning the confidentiality, integrity, and availability of sensitive medical data. Traditional security measures are often insufficient to address the unique challenges posed by IoT environments, such as heterogeneity, resource constraints, and the need for real-time processing. To tackle these challenges, we propose a comprehensive three-phase security framework designed to enhance the security and reliability of IoT-enabled healthcare systems. In the first phase, the framework assesses the reliability of IoT devices using a reputation-based trust estimation mechanism, which combines device behavior analytics with off-chain data storage to ensure scalability. The second phase integrates blockchain technology with a lightweight proof-of-work mechanism, ensuring data immutability, secure communication, and resistance to unauthorized access. The third phase employs a lightweight Long Short-Term Memory (LSTM) model for anomaly detection and classification, enabling real-time identification of cyber threats. Simulation results demonstrate that the proposed framework outperforms existing methods, achieving a 2% increase in precision, accuracy, and recall, a 5% higher attack detection rate, and a 3% reduction in false alarm rate. These improvements highlight the framework's ability to address critical security concerns while maintaining scalability and real-time performance.

✉ Behnam Rezaei Bezanjani
B.rezai@kmu.ac.ir

✉ Seyyed Hamid Ghafouri*
Ghafoori@iauk.ac.ir

✉ Reza Gholamrezaei
Gholamrezaei@iauk.ac.ir

[1]Department of Computer Engineering, Kerman Branch, Islamic Azad University, Kerman, Iran.

# 1 Introduction

The rapid advancement of IoT technology has led to its widespread adoption in various sectors, including healthcare. Integrating IoT devices in healthcare systems promises significant improvements in patient care and data management, enabling real-time monitoring, personalized treatments, and efficient management of medical records. However, this widespread use of IoT devices also brings critical security challenges, particularly in safeguarding sensitive medical data from cyber threats [1]. Healthcare data is inherently sensitive, necessitating robust measures to ensure its confidentiality, integrity, and availability. Traditional security solutions often fail to address the unique challenges posed by IoT in healthcare, such as the heterogeneity of devices, the vast amount of data generated, and the need for real-time processing and response [2-3]. Therefore, ensuring the security of IoT-based healthcare systems is a paramount concern, calling for innovative and comprehensive security frameworks.

Recent studies have investigated various methods to increase the security of IoT systems. The authors in [3] proposed an artificial rabbit optimizer coupled with deep learning models to secure smart healthcare systems using blockchain technology. Similarly, GTxChain, an intelligent IoT blockchain architecture based on graph neural networks, is introduced in [4], which provides a promising approach to address security concerns in IoT environments. These studies emphasize the importance of using advanced technologies such as blockchain and artificial intelligence (AI) to strengthen IoT security frameworks. With its decentralized and immutable nature, blockchain technology provides a robust platform for securing medical data, ensuring transparency, and preventing unauthorized access or tampering [5-6]. Integrating AI, particularly machine learning, further enhances the capability to detect and respond to sophisticated cyber threats by identifying patterns and anomalies that traditional methods might overlook [7-8].

Kumar et al. [36] highlighted the necessity of advanced cybersecurity strategies for combating emerging threats like deepfake disinformation, which has parallels in IoT-based healthcare security. Similarly, decentralized frameworks, such as those proposed in [37], demonstrate the potential for secure, scalable IoT systems.

Despite these advancements, there remains a need for comprehensive methods that combine these technologies to address the multifaceted security challenges in IoT-based healthcare systems. Hybrid approaches, such as those discussed by Kumar et al. [38-40], emphasize the importance of combining fuzzy decision-making, explainable AI, and blockchain for reliable and interpretable security frameworks. This paper aims to contribute to this growing body of knowledge by proposing a novel security framework that integrates trust management, blockchain, and AI to enhance the security and reliability of IoT-enabled healthcare environments.

## 1.1 Problem Statement

Some challenges in previous work have not been resolved. Among them are the inability to predict new attacks and the lack of generalizability of the methods used in the references.

- **Limitations of Conventional Machine Learning Techniques in Detecting Novel and Complex Threats (e.g., Adversarial Attacks and Zero-Day Attacks)**: Conventional machine learning (ML) techniques have been commonly applied in attack detection tasks due to their good performance accuracy. However, these techniques have been criticized for their limited ability to detect novel and complex threats. They often struggle to identify sophisticated and emerging attack patterns [1-4].
- **Limited Coverage of Attack Types in Previous Research References:** Many of these studies have focused on only one or two specific types of attacks, neglecting the broader spectrum of potential threats in cybersecurity. This limited coverage hinders the comprehensive understanding and detection of various attack patterns, leaving significant gaps in the security framework [10, 20, 22].
- **Scalability**: Many existing blockchain-based solutions struggle to handle the high volume of data generated by IoT devices in healthcare environments, leading to latency and resource constraints. This limitation poses significant challenges for real-time applications where timely responses are critical.
- **Interoperability**: IoT-driven healthcare systems often involve heterogeneous devices with varying computational capabilities and protocols. Existing solutions lack the flexibility to seamlessly integrate with such diverse ecosystems, leading to gaps in security coverage.
- **Resource Constraints**: Traditional intrusion detection systems often rely on computationally intensive algorithms, making them unsuitable for resource-constrained IoT devices typically used in healthcare applications.
- **Data Privacy and Security**: The sensitive nature of healthcare data demands robust measures to prevent unauthorized access and ensure regulatory compliance. However, many existing frameworks fail to balance privacy, security, and system performance effectively.

## 1.2 Contributions

Our main contributions are unfolded as follows:

- **Lightweight Proof-of-Work Algorithm**: To further improve the system's efficiency and security, a lightweight proof-of-work (PoW) algorithm is presented. This algorithm minimizes the computational requirements for participating nodes, making it feasible for resource-constrained IoT devices. The lightweight PoW algorithm maintains the integrity and security of the blockchain by preventing unauthorized access and ensuring data authenticity without imposing significant computational burdens on the devices.
- **Presenting a lightweight LSTM model**: The Long Short-Term Memory (LSTM) is a lightweight model designed to be computationally efficient. This approach reduces training and testing time, making it suitable for real-time applications in IoT-based healthcare environments.

- Improved Detection Performance: Simulation results show that the proposed framework achieves a 2% increase in precision, accuracy, and recall, a 5% higher attack detection rate, and a 3% reduction in false alarm rate compared to existing methods. These improvements demonstrate the framework's ability to address critical security concerns effectively.
- **Enhancing medical data security**: The proposed method preserves data privacy well using an innovative combination of BC technology and bi-directional short-term memory networks.
- **Predicting more than 10 attack types:** We evaluated the proposed method in two different datasets and compared it with three recent methods, showing its superiority.

Some abbreviations are used in the text. Table 1 shows the description of these abbreviations.

**Table 1.** Description of the abbreviations.

| Symbol | Description | Symbol | Description |
|---|---|---|---|
| IoT | Internet of Things | IPFS | InterPlanetary File System |
| IoV | Internet of Vehicles | DH | Distributed hash table |
| CNN | Convolutional Neural Networks | CA | Content addressed hash value |
| CT | Confident threshold | CPS | Cyber Physical System |
| TV | Transaction value | LSTM | Long Short-Term Memory |
| TS | Trust score | NT | Number of Transaction |
| BC | Blockchain | BF | BlockFog |

### 1.3 Paper Organization

The subsequent sections will be structured as follows: Section II will assess and analyze previous studies and relevant research. Section III outlines the proposed method, including its steps and specific details. Section IV presents a thorough analysis of the simulation results. Finally, Section V presents the future works and conclusion of the research paper.

## 2 Related Work

The integration of advanced security frameworks in IoT-based systems has been extensively studied in recent literature, highlighting the significance of secure communication protocols and decentralized technologies. Kumar and Mishra [41] proposed a fuzzy-based computational technique to assess vulnerabilities in urban systems, emphasizing adaptability and scalability—concepts that align well with IoT healthcare security. Similarly, Kumar et al. [42] and Kumar and Ahmad Khan [44] explored blockchain-based solutions for securing communication protocols in military and critical systems, providing insights into the robustness and reliability of decentralized approaches. These studies underline the importance of ensuring data integrity and preventing unauthorized access in sensitive environments, which are pivotal in healthcare IoT frameworks. Furthermore, Pandey et al. [43] presented a secure cyber-engineering framework tailored for IoT-enabled smart healthcare systems, incorporating blockchain and AI technologies to enhance system resilience against sophisticated threats. Collectively, these works provide a foundation for designing scalable, secure, and efficient frameworks, addressing challenges pertinent to IoT-based healthcare environments.

A comprehensive review of integrating blockchain and deep learning in various applications has been conducted in [11]. The review highlighted the synergistic potential of these technologies in enhancing data security, privacy, and decentralized learning frameworks. It identified key trends and future research directions and emphasized developing scalable and efficient solutions.

[13] proposes multi-layer deep learning approaches to identifying botnet attacks in IoT industrial systems. Their approach combines convolutional neural networks (CNN) and short-term memory (LSTM) networks to analyze network traffic data. The proposed method achieves high accuracy and low false positive rates and effectively increases cyber security measures.

In [14], the authors present BlockMedCare, a healthcare system that uses IoT, blockchain, and IPFS for secure data management. This system ensures data integrity, privacy, and availability in healthcare applications. Their results showed improved data security and management efficiency compared to traditional systems.

An energy-efficient encryption and authentication technique designed for IoT security is presented in [15]. The proposed method uses lightweight cryptographic algorithms to reduce computational overhead while maintaining strong security. Experimental results showed significant improvements in energy efficiency and processing time.

[16], have focused on anomaly detection in IoT time series data using a convolutional iterative autoencoder. Their model combined convolutional and recurrent neural networks to capture spatial and temporal patterns in the data. The proposed method achieved high detection accuracy and effectively-identified anomalies in complex IoT datasets.

[17] proposes a blockchain-based authentication and key agreement protocol for the Internet of Vehicles (IoV). This protocol aims to increase the security and efficiency of vehicular communication by using blockchain for secure key management. Their results show better authentication speed and security compared to existing methods.

A secure fusion approach for multi-robot systems is developed in [18]. This approach uses blockchain to ensure data integrity and coordination between independent entities. Experimental results showed increased data security and system performance in intelligent autonomous systems.

In [19], the authors addressed optimizing dynamic data storage in the IoT using blockchain for wireless sensor networks. The optimized method of dynamic data storage (ODSD) is proposed to improve the efficiency and security of data management. The results indicated significant improvements in data recovery speed and storage optimization.

In [20], the use of blockchain technology for secure communication of health care data in IoT architecture in 5G networks is investigated. Their approach uses blockchain to ensure privacy and data integrity among non-terminal nodes. The proposed method showed improved security and efficiency in healthcare data transmission.

[21] proposes a distributed network security framework for the Internet of Energy based on the IoT. This framework aims to increase the flexibility and security of energy systems against cyber threats using blockchain technology. The results showed that network security and system strength have improved.

Committee consensus introduced a decentralized, federated learning framework [54] using blockchain [23]. This framework aimed to increase data privacy and collaborative learning by decentralizing the model training process. The experimental results showed that the model accuracy and data privacy were improved compared to centralized approaches.

In [24], a secure evidence and incident management framework for vehicular networks using deep learning and blockchain is proposed. This framework aims to increase the reliability and security of incident-handling processes. The results showed an improvement in accident detection accuracy and data integrity.

In [25], he developed outlier models to secure data migration in cloud centers. Their approach used mixed localization-based models to identify and reduce outliers during data migration. Experimental results indicate increased data integrity and security during cloud-based operations.

The authors in [26] have presented a blockchain-based data protection framework for modern power systems to protect against cyber-attacks. Their proposed method uses blockchain to increase data security and system flexibility. The results show a significant improvement in data protection and system robustness.

[27] proposes an elastic and cost-effective data carrier architecture for smart contracts on the blockchain. This architecture addresses scalability and efficiency issues in the implementation of smart contracts. Experimental results show performance improvement and cost reduction compared to traditional architectures.

The authors presented a semi-supervised learning framework for detecting distributed attacks in [28]. Their approach combined supervised and unsupervised learning techniques to increase recognition accuracy. The results indicated an improvement in the attack detection rate and a decrease in false positives.

[29], have investigated intrusion detection based on deep learning with adversarial learning. Their approach used adversarial training to increase the robustness of intrusion detection systems. The experimental results showed a significant improvement in the detection accuracy and flexibility of the system against hostile attacks.

These studies highlight ongoing advances and challenges in integrating blockchain, deep learning, and the IoT and emphasize the importance of developing scalable, secure, and efficient solutions for emerging technology ecosystems.

Recent advancements in integrating blockchain, artificial intelligence, and other computational techniques have paved the way for improved data security and privacy in healthcare and other domains. Ch (2024) proposed a real-time communication framework for blood bank systems using XGBoost and Supabase, emphasizing secure access control in time-critical applications [45]. Similarly, Sree et al. (2024) introduced a fuzzy logic-based cipher mechanism to mitigate passive attacks, providing a novel approach for robust data encryption [46]. Privacy-driven applications have also gained traction, as demonstrated by Guggilam et al. (2024), who utilized YOLOv8 and SHA-256 to enhance vessel detection while ensuring privacy in sensitive maritime operations [47].

In the healthcare domain, Rupa et al. (2024) developed a hash-based DCIWT approach for tampering detection in IoMT (Internet of Medical Things), ensuring privacy preservation in medical data [48]. Furthermore, the use of blockchain technology for privacy-centric decentralized applications (DApps) was explored by Rupa et al. (2022), showcasing a knowledge engineering framework for protecting medical certificate privacy [49]. These works underline the significance of combining blockchain and AI-driven techniques to address data security challenges in various domains, including healthcare, which aligns with the focus of the proposed framework.
An analytical comparison of the work done is shown in Table 2.

**Table 2.** Comparative analysis of different approaches.

| Ref. | Year | Advantages | Disadvantages |
|---|---|---|---|
| [12] | 2023 | Detailed taxonomy and identification of challenges; useful for guiding future research. | Lacks specific solutions or case studies addressing the challenges identified |
| [13] | 2022 | High accuracy and low false positive rates in botnet attack detection. | Focused on a specific type of attack, which may limit generalizability to other security threats. |
| [14] | 2022 | Improved data security and management efficiency in healthcare applications. | Implementation complexity and potential scalability issues in large-scale systems. |
| [16] | 2020 | High detection accuracy for anomalies in IoT time series data. | It is computationally intensive, which may limit real-time application. |
| [17] | 2021 | Enhanced security and efficiency in vehicular communications. | Potentially high implementation costs and complexity. |
| [18] | 2021 | Improved data security and system performance in multi-robot systems. | Limited to smart autonomous systems; scalability concerns. |
| [19] | 2021 | Enhanced data retrieval speed and storage optimization in IoT. | Specific to wireless sensor networks, limiting broader applicability. |
| [21] | 2021 | Improved network security and system robustness for the energy internet. | Potentially high implementation costs and integration challenges. |
| [25] | 2019 | Enhanced data integrity and security during cloud-based data migration. | Potentially high computational requirements for large datasets. |
| [26] | 2018 | Improved performance and reduced costs in smart contract execution. | Focused on power systems, which may limit applicability to other domains. |

| Ref. | Year | Advantages | Disadvantages |
|---|---|---|---|
| [29] | 2018 | Significant improvements in detection accuracy and system resilience against adversarial attacks. | Adversarial training can be computationally intensive. |

Table 3 compares the proposed method with other existing methods and highlights the advantages of the proposed framework.

**Table 3.** Table 3. Comparative Analysis of Proposed Method and Existing Approaches.

| Advantage | Proposed Method | Ref. [3] | Ref. [4] | Ref. [37] | Ref. [43] |
|---|---|---|---|---|---|
| Scalability | ✓ | ✕ | ✕ | ✓ | ✕ |
| Lightweight Design for Resource-Constrained IoT | ✓ | ✕ | ✕ | ✕ | ✕ |
| Real-Time Anomaly Detection | ✓ | ✕ | ✕ | ✕ | ✓ |
| Trust Management for Device Reliability | ✓ | ✕ | ✕ | ✕ | ✕ |
| Blockchain-Based Data Immutability | ✓ | ✓ | ✓ | ✓ | ✓ |
| Integration of AI for Cyber Threat Detection | ✓ | ✕ | ✓ | ✕ | ✓ |
| Evaluation Across Comprehensive Metrics | ✓ | ✕ | ✕ | ✕ | ✕ |
| Multi-Layer Security (Blockchain, AI, Trust) | ✓ | ✕ | ✕ | ✕ | ✕ |

# 3 Proposed Method

This section explains the proposed method in detail. The proposed methodology integrates trust management, blockchain-based privacy preservation, and anomaly detection using AI to enhance the security of IoT-based healthcare systems. In the trust management phase, a reputation-based trust estimation model is employed to calculate the trustworthiness of IoT devices. This model uses parameters such as transaction value, frequency, and behavior patterns to categorize transactions as valid, suspicious, or malicious. The categorization process is grounded in statistical thresholds derived from historical data. To enhance this phase, a robust statistical approach is introduced to validate the trust score thresholds. These thresholds are dynamically updated using adaptive algorithms that analyze real-time data patterns, ensuring the model remains accurate and responsive. The blockchain-based privacy preservation phase utilizes a lightweight Proof-of-Work (PoW) algorithm to address the resource constraints typical in IoT environments. This algorithm reduces computational overhead while maintaining data authenticity and integrity. Unlike traditional PoW, which requires intensive hash computations, the lightweight PoW optimizes the process with simplified cryptographic puzzles that are computationally feasible for IoT devices.

Additionally, sensitive data is stored off-chain using the InterPlanetary File System (IPFS), which enhances scalability and data security while reducing the storage burden on the blockchain. This approach also facilitates rapid data retrieval and efficient management.

In the AI-driven anomaly detection phase, a lightweight Long Short-Term Memory (LSTM) model is used to capture temporal dependencies in IoT data streams, enabling the detection of sophisticated attack patterns. To further optimize this process, Singular Value Decomposition (SVD) is applied, reducing the computational complexity of the LSTM model. Complementing this is the use of a Variational Autoencoder (VAE), which transforms sensitive data into encoded representations to mitigate risks of inference attacks. Unlike traditional VAEs, this implementation ensures high fidelity in data reconstruction and includes an added noise layer to enhance security further. The flowchart of the proposed method is shown in Figure 1.

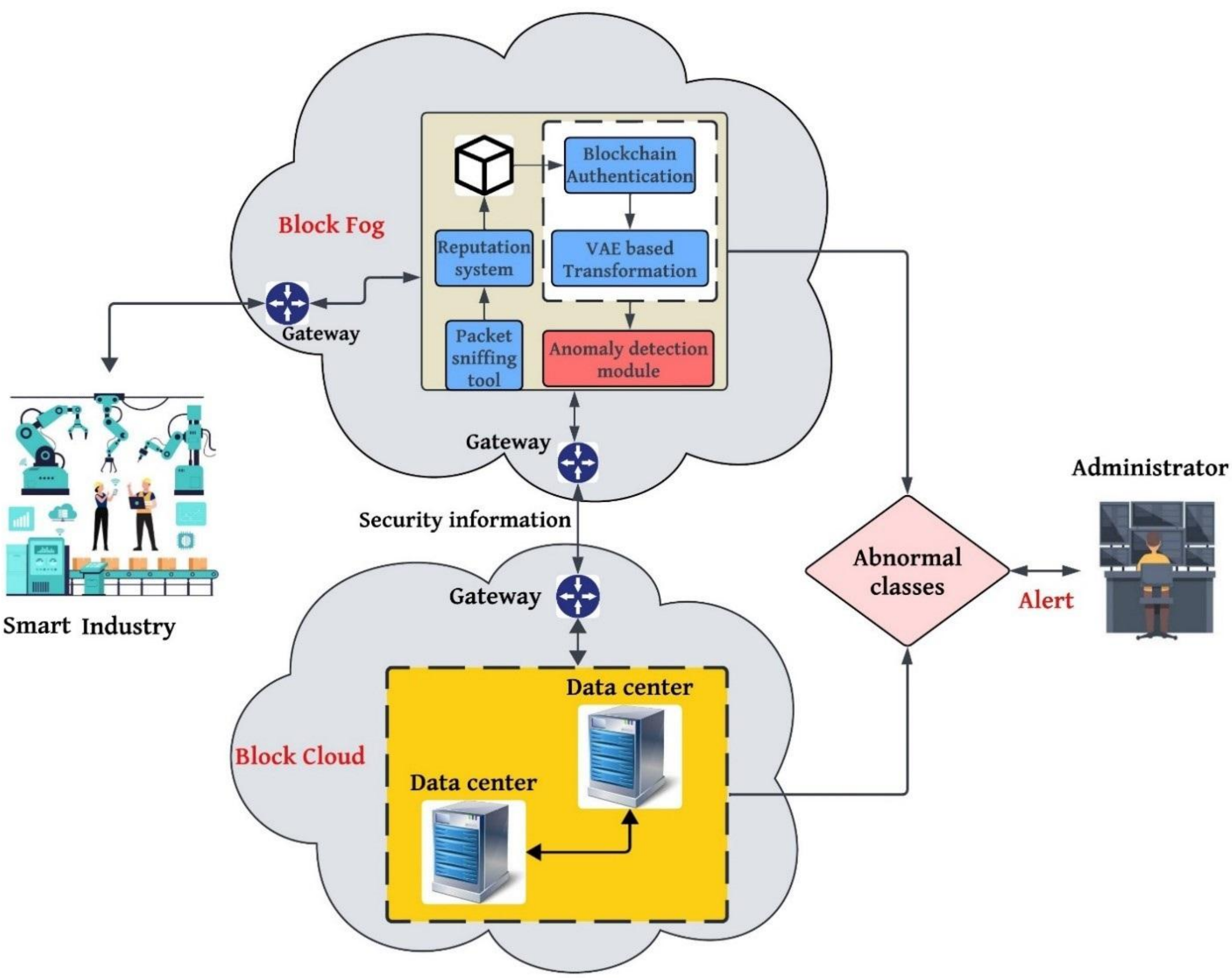

**Fig. 1**. Flowchart of the proposed method.

The phases of the proposed framework are tightly integrated to ensure seamless and efficient operation. The process begins with trust scores, which serve as an initial filter for device transactions. Only transactions that meet the acceptable trust thresholds are allowed to proceed to the blockchain authentication phase. The blockchain's lightweight Proof-of-Work (PoW) algorithm then ensures that authenticated data is securely logged and prepared for further analysis. This authenticated data is subsequently processed by the AI-driven anomaly detection module, where sophisticated techniques identify potential threats or irregularities. Any detected anomalies are fed back into the trust estimation phase, enabling the system to adapt and improve its resilience dynamically over time. This cyclical integration creates a robust and responsive cybersecurity framework.

## 3.1 Phase 1. Trust management

Today, attackers carry out different types of attacks, which include active (data poisoning attacks) and passive (discovery of publicly available data by a hacker). Data poisoning attacks are carried out by altering normal data, and injecting false data is a prime example of these attacks. For this reason, protecting data integrity is one of the requirements of physical cyber systems. There are three layers for privacy protection in the proposed architecture: device layer, network layer, and application layer. Data viewed by IoT devices is generated using the device layer. Then, to ensure the security of the data, their correctness is verified in parallel. Transaction value, threshold value (TV), and trust score (TS) are the parameters involved in the estimation model. Reputation score estimation uses the number of transactions calculated based on the TS value. Then, the extracted features are classified based on malicious, reliable, and valid transactions.

The reliability of the blockchain framework is important due to the existence of a malicious entity that always manipulates information. The addressable hash value is generated by the transaction sensor information, which is the output of the off-chain data storage, which results in the initialization of the addressable hash value parameters and the number of transactions to zero. Then, in the distributed hash table, the information is stored by the sensor devices and transferred to the BlockFog and BlockCloud. There is a set of fog nodes $FN = \{FN_1.FN_2.\dots.FN_N\}$ in the network layer. The communication of each node with other nodes is based on the peer-to-peer mode by sensor nodes $SN = \{SN_1.SN_2.\dots.SN_N\}$.

Trust value estimation is done using the transaction's trust score. The data sets minimum and maximum trust threshold values are compared with each transaction to identify the trusted devices with the same sensor devices with higher trust values. The entire process mentioned above is shown in Algorithm 1, based on which malicious, good, and normal transactions are classified.

| **Algorithm 1**. Reputation-based Trust Estimation. |
| --- |
| **1. Procedure** ESTIMATE_REPUTATION (TV) |
| **2. Initialize** the (TS) = 0; |
| **3.** Estimate CT range (min ($x_i$), max ($x_i$)), where x ∈ X; |
| **4**. Read TV of the sensor data obtained from the dataset (X); |
| **5**. Based on the TV, the TS and CT are estimated; |
| **6. if** TV = CT **then**: |
| 7. $TS^+$ = 1; |
| **8. else** |
| **9**. $TS^-$ = 0; |
| **10. end if** |
| **11**. Consequently, estimate the reputation score as follows: |

12. TS $=\frac{\frac{TS}{10}}{NT}$;

13. Then, the transaction is categorized according to the estimated TS and the total number of features $(F) = \{f_1.f_2...f_n\}$ in dataset X;

**14. if** $(TS > \frac{FC-2}{10}$ && $TS \leq \frac{FC}{10})$ **then:**

15. "Valid Transaction";

**16. else if** $(TS \geq \frac{FC-5}{10}$ && $TS \leq \frac{FC-2}{10})$ **then**:

17. "Reliable Transaction";

**18. else:**

19. "Malevolent Transaction";

**20. end if**

**21. end procedure**

After the transactions are sorted, the data is stored off-chain for the information to be stored in IPFS. The reliability of the sensor device is based on reputation. To avoid duplicate data, a unique addressable hash value is generated. Algorithm 2 shows how to do this.

**Algorithm 2**. Off-Chain Data Storage.

1. **Input**: Information of transaction sensor;
2. **Output**: Content addressable hash value;
3. **Initialization**
4. At first, the content addressed the hash value CA, and the number of transactions NT are initialized as 0;
5. CA = 0;
6. NT = 0;
7. **for** all sensor devices, **do:**
8. NT→ Obtain values from the sensor device;
9. **if** NT is valid, **then**:
10. CA Estimate the content addressed to the hash value;
11. Store the updated CA into the IPFS of sensors;
12. The IPFS stores the information in the DHT;
13. The addressable hash is transmitted to both BF and BC;
14. **end if**
15. **end for**

## 3.2 Phase 2. Blockchain-based privacy-preserving

First, a lightweight proof-of-work algorithm based on a blockchain has been used for data authentication. Then, the encoding format transforms the features at the next level to reduce the inference attacks. These mechanisms focus on data authentication, transformation, and model generation. The CPS network uses a method that converts data into a secure format with asymmetric encryption to protect against unauthorized access and tampering. This process involves creating a specially encrypted digest of the data to ensure that it remains unchanged. Even a small change in the data results in a completely different summary, making it easy to spot any manipulation. This encrypted digest is used along with other information to create a block on the blockchain. Any change in the data affects the entire chain of summaries, which is easily reviewed. Blockchain ensures data integrity through PoW, which typically requires significant computing power. However, the proposed method is less demanding regarding computing resources to maintain data integrity. When data needs to be verified in a blockchain, the authentication algorithm processes various inputs such as transaction numbers, previous and current hash values, timestamps, and proofs. The data generates a hash value for the block if the data is valid. This process is repeated for each block, and after confirmation, the new block is added to the blockchain. The blockchain design ensures that it cannot be altered once data is verified and maintains its integrity even on untrusted networks.

### 3.2.1 Privacy-Based Variational Auto-Encryptor

A Variational Autoencoder (VAE) is an advanced type of autoencoder designed to generate new samples from data based on its underlying distribution. This capability enhances systems for anomaly detection. Unlike traditional autoencoders and denoising models, which only learn and replicate hidden features, VAEs can create novel data examples. Training a VAE involves encoding an input dataset, which lacks class labels, into a latent representation using adjusted weights. These latent codes are then reconstructed into the original data format through a set of generative weights derived from the latent representation. The training process involves a dataset with multiple features and numerous records, where a hidden variable introduces randomness. The VAE's stochastic method includes defining prior distributions and calculating probabilities to enable the generation of new data points. The prior distribution $p_{\mu}(Q)$ is used to estimate hidden variables $Q_i$. After estimating these hidden variables, the observed data point $x_i$ is generated based on the conditional probability distribution $p_{\mu}(x|Q)$. Essentially, $p_{\mu}(Q)$ acts as the prior distribution while $p_{\mu}(x|Q)$ is the conditional distribution determined by the general parametric relationships between these two distributions. To find the marginal probability of a feature X, you sum the probabilities of each data point, following the method outlined in (1) and (2).

$$log\ p_{\mu}\left(x_1,x_2,...,x_N\right) \ = \ \sum_{i=1}^{N}\log p_{\mu}\left(x_i\right) \tag{1}$$

$$log\ p_{\mu}\left(x_i\right) = D_C(q_{\Gamma}(Q|x_i)\ K\ p_{\mu}(Q|x_i)) + L(\mu,\Gamma;x_i) \tag{2}$$

In (1), $D_C(q_{\Gamma}(Q|x_i)\ K\ p_{\mu}(Q|x_i))$ represents the Coleback-Leibler divergence between the estimated prior $q_{\Gamma}(Q|x_i)$ and the true prior $p_{\mu}(Q|x_i)$, and its value is a non-negative value estimated by (3).

$$D_{KL}(q_{\varphi}\|\ p_{\theta}) = \sum_{i=1}^{N} p_{\theta}(x_i)\log(\frac{p_{\theta}(x_i)}{q_{\theta}(x_i)}) \tag{3}$$

$L(\mu,\Gamma;x_i)$ refers to the variable lower bound on the final probability for each data point i, as shown in (4).

$$L(\mu,\Gamma;x_i) = -D_C(q_\Gamma(\mathrm{Q}|\mathrm{x}_i) \| p_\mu(Q)) + E_{q_\Gamma(\mu|\mathrm{x}_i)}[\log(p_\mu(x_i \mid Q))] \tag{4}$$

The model fits $q_\Gamma(Q \mid x_i)$ using a normal distribution that approximates the true input distribution. Leveraging built-in hidden distributions minimizes the likelihood of alterations in the original input. Optimization is achieved through an L2 regularization function, transforming the original data into a new format. As a result, the generated data is robust against inference attacks since the model can produce new samples from various potential distributions rather than relying on the original data alone.

## 3.3 Phase 3. Anomaly detection module based on lightweight LSTM

Recurrent Neural Networks (RNNs) are deep learning algorithms designed for classifying sequential data. They extend conventional feedforward neural networks by incorporating recurrent connections for improved modeling. In the context of smart city data classification, a lightweight Long-Short-Term Memory (LSTM) model is developed to address the limited capacity of IoT devices, reducing complexity. LSTMs are chosen over traditional RNNs because RNNs, while popular for time series forecasting, struggle with long-term dependencies. LSTMs, with their unique architecture consisting of forgetting, input, and output gates, effectively mitigate these limitations. Figure 2 illustrates a typical LSTM cell structure, showcasing these key components. The operations of the input, forget, and output gates, as well as the hidden vector, are expressed through element-wise multiplication and the summation of matrices, as outlined in the following equations:

$$i_t = sigmoid(W_{ii} p_t + W_{hi} q_{t-1} + b_i) \tag{5}$$

$$d_t = sigmoid(W_{if} p_t + W_{hf} q_{t-1} + b_f) \tag{6}$$

$$g_t = tanh(W_{ig} p_t + W_{hg} q_{t-1} + b_g) \tag{7}$$

$$o_t = sigmoid(W_{io} p_t + W_{ho} q_{t-1} + b_o) \tag{8}$$

$$c_t = d_t * c_{t-1} + i_t * g_t \tag{9}$$

$$q_t = o_t * tanh(c_t) \tag{10}$$

$$\begin{pmatrix} i_t \\ d_t \\ g_t \\ o_t \end{pmatrix} = \begin{pmatrix} sigmoid \\ sigmoid \\ \tanh \\ sigmoid \end{pmatrix} * (W_i p_t + W_h q_{t-1} + B) \tag{11}$$

$$W_i = [W_{ii}, W_{if}, W_{ig}, W_{io}]^T \quad (12)$$

$$W_h = [W_{hi}, W_{hf}, W_{hg}, W_{ho}]^T \quad (13)$$

$$B = [b_i, b_f, b_g, b_o]^T \quad (14)$$

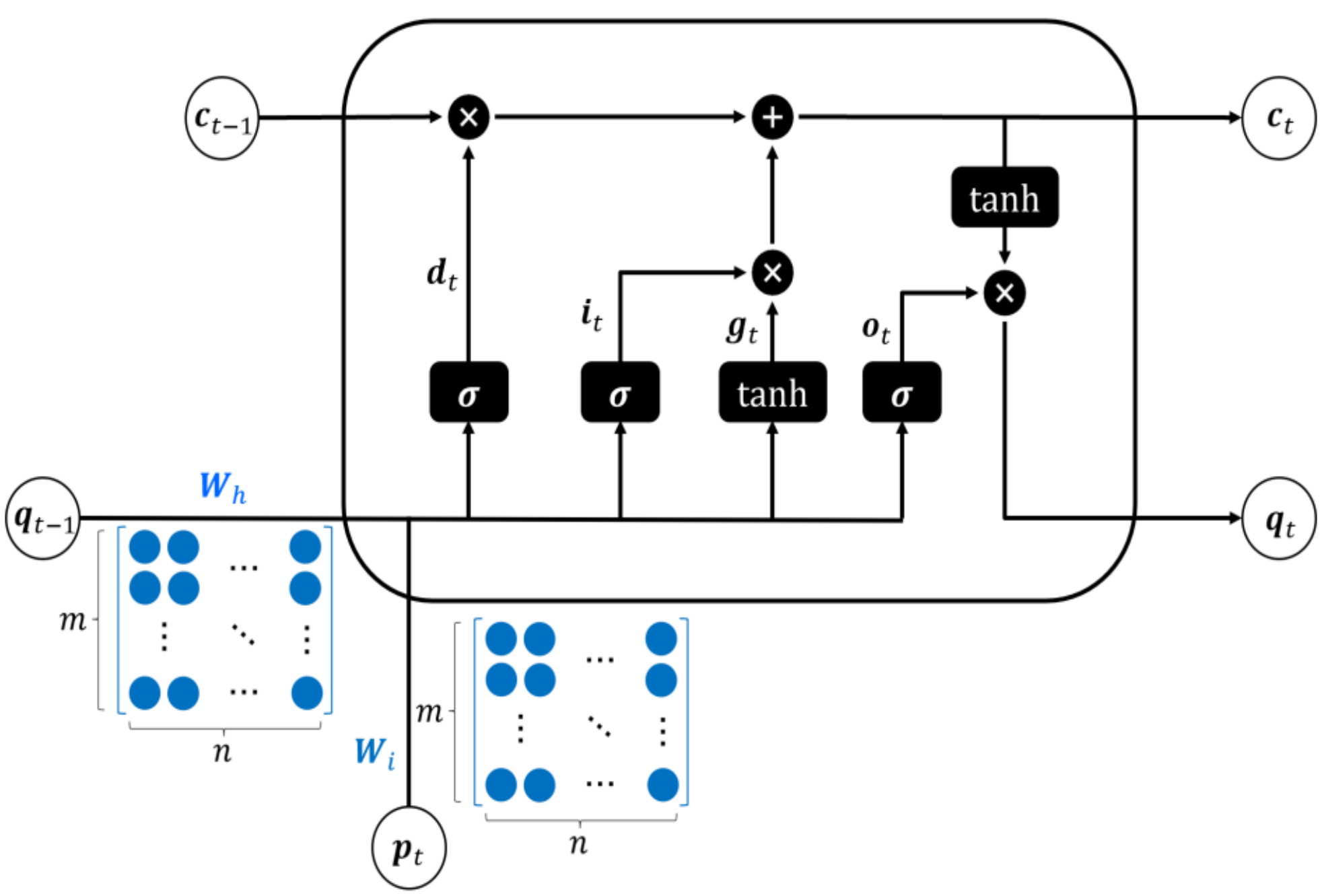

**Fig. 2**. LSTM network architecture.

The typical LSTM model in IoT devices is often too large or complex for practical deployment. Most of the computational complexity in conventional LSTM models arises from the matrix-vector multiplications, specifically $W_i\, p_t$ and $W_h\, q_{t-1}$. To mitigate this, singular value decomposition (SVD) can be employed to reduce the computational load. In SVD, matrix $W \in R^{m \times n}$ is decomposed into U, $\sum$, and V matrices, where U has dimensions m*r, S is a diagonal matrix containing singular values, and V has dimensions r*n. The dimensions of matrices U and V can be reduced by truncating the singular values and setting smaller ones to zero. This reduction involves eliminating columns corresponding to these small singular values. Thus, by choosing an appropriate threshold r (where $r < min(m.n)$), matrix W can be approximated as described in equation (15).

$$\begin{aligned} & W_{m\times n} = U_{m\times m} \Sigma_{m\times n} V_{n\times n}^T \approx \\ & U_{m\times r} \Sigma_{r\times r} V_{r\times n}^T \approx \\ & U_{m\times r} N_{r\times n} \end{aligned} \quad (15)$$

In the proposed model, the matrix W is decomposed into two matrices, $W_1$ and $W_2$, where W has dimensions m × n, $W_1$ has dimensions m × r, and $W_2$ has dimensions r × n. In traditional LSTM models, we typically require m × n weights for W, whereas in the proposed approach, we only need (m × r + r × n) weights for $W_1$ and $W_2$ combined. Figure 3 shows the structure of the proposed style LSTM by applying dimension reduction to the matrices $W_h$ and $W_i$ the regular LSTM model.

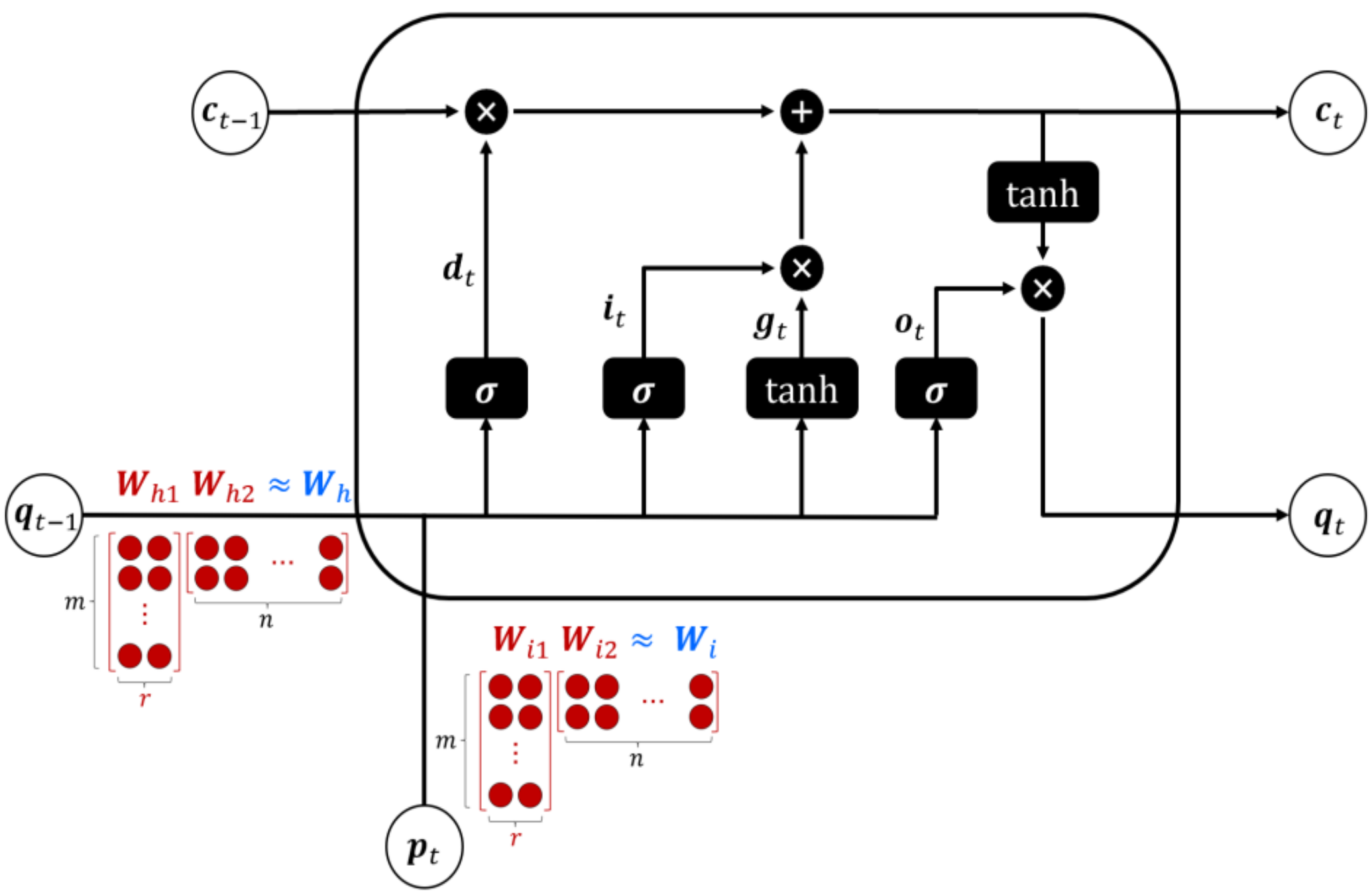

**Fig. 3**. Proposed lightweight LSTM cell structure.

# 4 Simulation Results

The proposed method is evaluated on two distinct datasets, NSL-KDD [31] and UNSW-NB15 [32]. Subsequently, the obtained results are compared with findings from prior works referenced as [10], [33], and [34]. This comparative analysis aims to gauge the effectiveness and performance of the proposed method compared to existing approaches on both UWSN and NSL-KDD datasets. Details of the dataset used in the simulation are given in Table 4 [21, 23, 24, 26, 28].

**Table 4.** Details of datasets NSL-KDD and UNSW-NB15.

| Name | Year | Records | Number of Features | Attacks |
|---|---|---|---|---|
| NSL-KDD | 1999 | 4,898,430 | 41 | DoS, Probe, U2R, R2L |
| UNSW-NB15 | 2015 | 2,540,044 | 49 | Backdoor, Worms, Shellcode, Exploits, Fuzzers, DoS, Generic, Reconnaissance, Analysis |

The simulations were performed on a system equipped with an Intel Core i7 processor, 32GB of RAM, and an NVIDIA RTX 4070 GPU. The software environment utilized Python 3.8, with TensorFlow for model training and NumPy for numerical computations. Additionally, blockchain functionality was simulated using the Hyperledger Fabric framework. For training, the models were run for 40 epochs with a batch size of 64, using the Adam optimizer with a learning rate of 0.001. Early stopping was applied to prevent overfitting.

## 4.1 Performance metrics

**Precision**: measures how well a model correctly identifies positive cases without including false positives (13).

$$Prec. = \frac{TP}{FP + TP} \tag{13}$$

**Recall**: quantifies how well a model identifies positive cases without missing any true positives. (14).

$$Rec. = \frac{TP}{FP + FN} \tag{14}$$

**F1-Score**: provides a comprehensive evaluation of a model's performance regarding positive predictions and capturing all relevant positive instances (15).

$$F1 - Score = \frac{2 \times Prec. \times Rec.}{Prec. + Rec.} \tag{15}$$

**Accuracy**: quantifies the overall correctness of predictions, which is calculated by (16) [35].

$$Accuracy = \frac{TP + TN}{Total\ Instances} \tag{16}$$

**Detection rate**: The metric assesses the percentage of positive cases accurately detected by a system (17).

$$DR = \frac{TP}{TP + FN} \tag{17}$$

**False alarm rate:** measures the rate at which the system produces false positive predictions or detections (18).

$$FAR = \frac{FP}{FP + TN} \tag{18}$$

## 4.2 Analysis of Results

Figure 4 depicts the training and validation accuracy changes and training and validation loss for two distinct datasets: NSL-KDD and UNSW-NB15. In NSL-KDD dataset, the training accuracy achieved 99%, and the validation accuracy reached 97% after 40 epochs. The choice of 40 epochs was guided by empirical analysis balancing performance and overfitting risk. Early stopping was employed as a strategy to halt training once the validation accuracy ceased to improve, preventing overfitting and unnecessary computational effort. This plateau suggests that the model effectively captured the underlying data patterns within this range, making further training redundant.

Training accuracy begins at 86% and steadily rises with each epoch, reaching 97% by epoch 40. Similarly, validation accuracy starts at 88% and exhibits a comparable upward trajectory, achieving 99.5% by epoch 40. Both metrics consistently improve throughout training, indicating effective learning and generalization by the LSTM network on the NSL-KDD dataset. Initially high, both training and validation losses decrease progressively with each epoch. This decline signifies the LSTM network's ability to reduce the disparity between predicted and actual values, demonstrating ongoing improvement in predictive performance on the NSL-KDD dataset. Training accuracy initiates at 83% and incrementally climbs to 95% by epoch 40, while validation accuracy starts at 84% and follows a similar upward trend, reaching 97.5% by the 40th epoch. Although both metrics show improvement, the enhancement rate slows after epoch 25, suggesting that the LSTM network effectively learns patterns in the UNSW-NB15 dataset but with diminishing returns over time. Both training and validation losses begin relatively high and decrease gradually over successive epochs. This reduction indicates the LSTM network's efficacy in minimizing errors between predicted and actual values on the UNSW-NB15 dataset. However, akin to accuracy trends, the loss reduction rate decelerates after epoch 25.

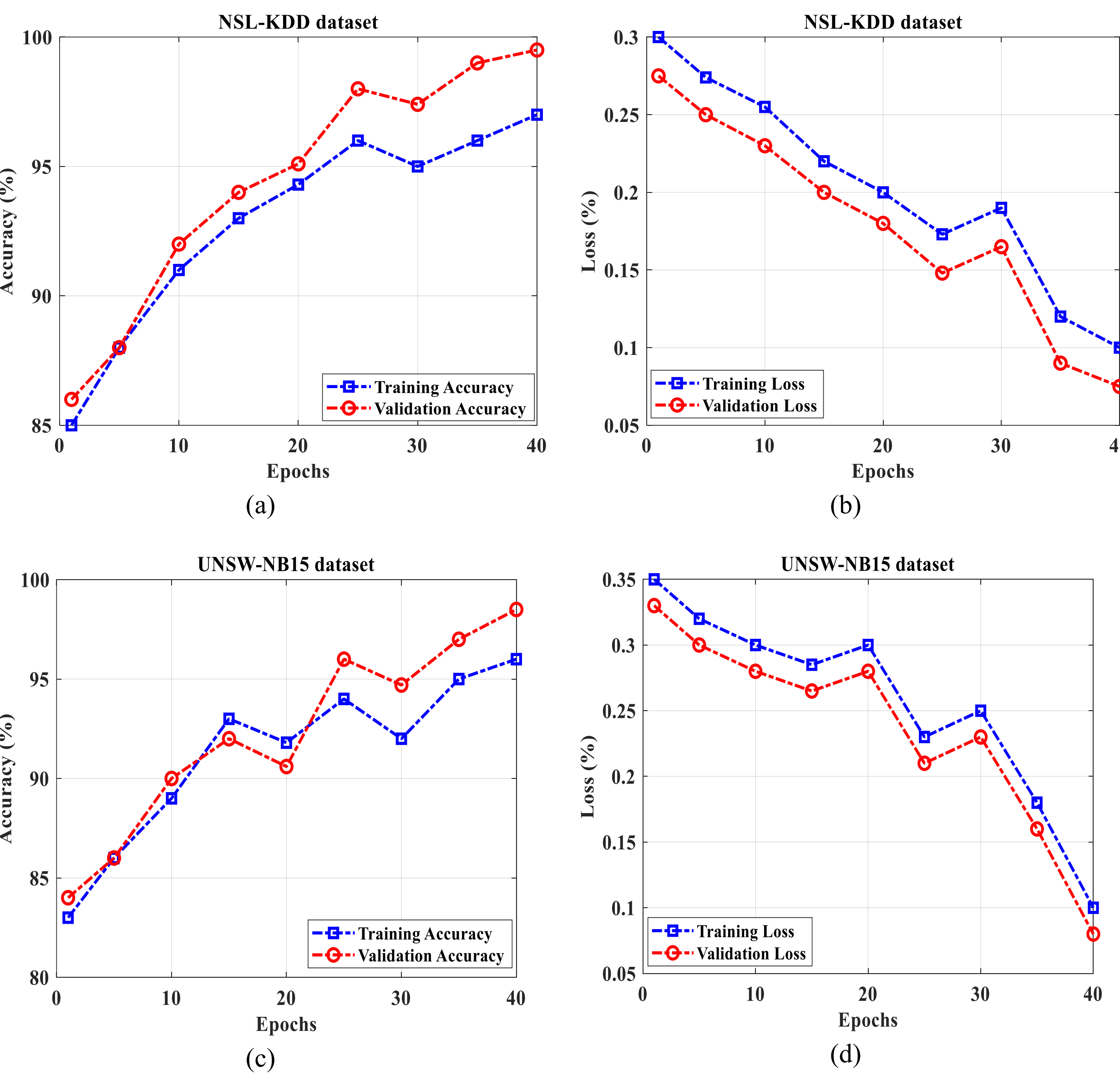

**Fig. 4** Training and validation performance on NSL-KDD and UNSW-NB15 datasets.

The NSL-KDD dataset exhibited a validation accuracy of 97%, compared to 95% for UNSW-NB15. These differences stem from the inherent characteristics of the datasets. NSL-KDD, being less complex, has fewer diverse attack patterns and a lower proportion of noise compared to UNSW-NB15, which encompasses a broader spectrum of modern attack scenarios and a higher level of overlap between normal and attack traffic features. These complexities in UNSW-NB15 lead to slight reductions in validation accuracy. Figure 4(b) demonstrates a similar reduction in loss for both datasets; however, the final loss values differ significantly (0.05 for NSL-KDD vs. 0.08 for UNSW-NB15). This disparity arises from the more intricate feature distributions in UNSW-NB15, which require additional processing to achieve comparable optimization. The higher noise levels and diverse attack types in UNSW-NB15 make it challenging for the model to converge to lower loss values despite effective learning dynamics. Figure 5, Table 3, and Table 4 compare the proposed method and other methods in terms of precision, recall, accuracy, and F1-score on the NSL-KDD dataset.

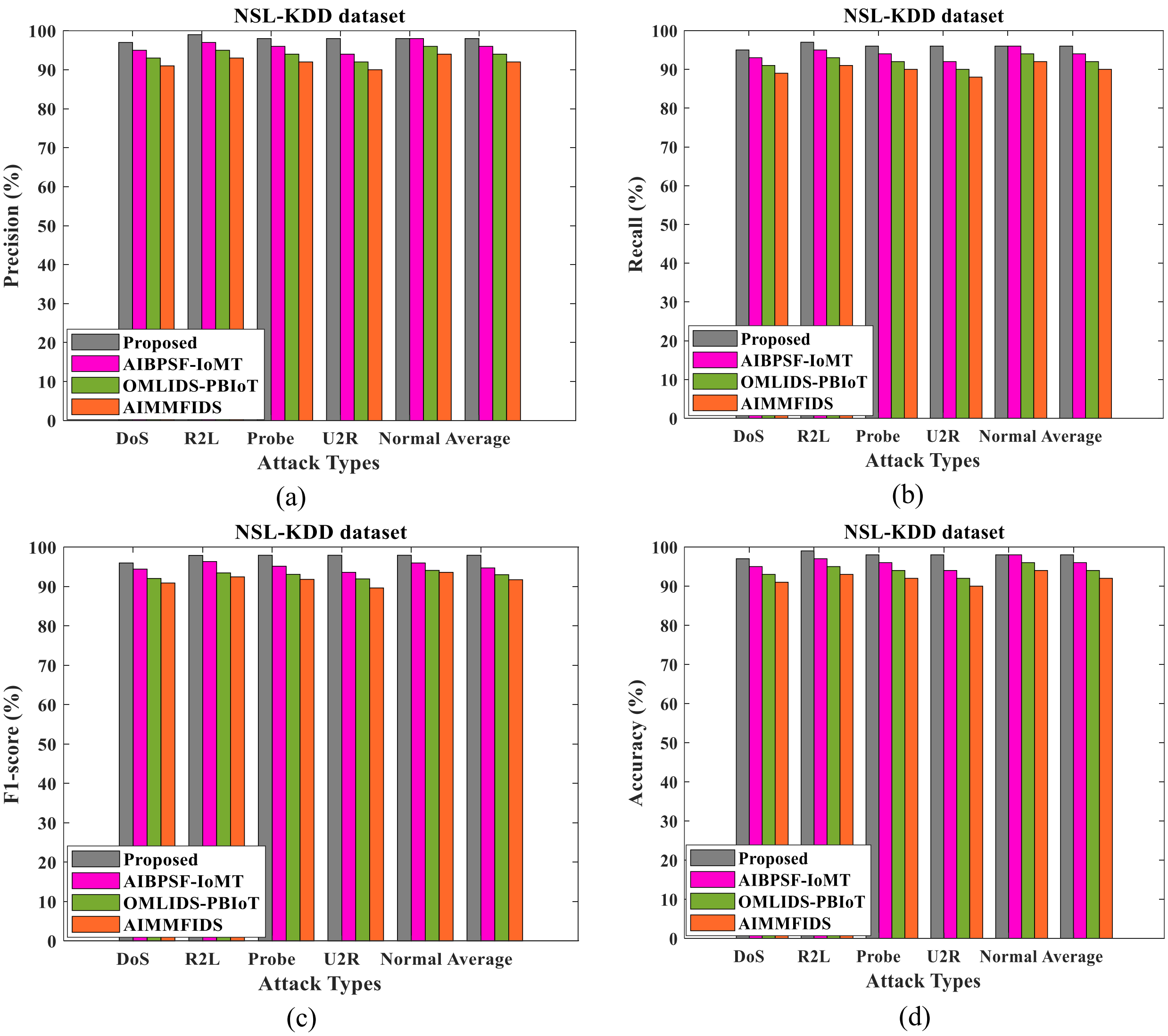

**Fig. 5** Comparative evaluation of the proposed method and three other methods in terms of precision, recall, F1-score, and accuracy on the NSL-KDD dataset.

**Precision, Recall, and F1-score:** As shown in the figure, the proposed method and other methods have performed better on the NSL-KDD dataset than the UNSW-NB15 dataset. This is due to the high number of attacks in the UNSW-NB15 data set, which is more complex than the NSL-KDD data. In addition, the LSTM network has identified attack patterns better than other methods due to the bidirectional nature of the calculations.

**Accuracy:** The LSTM network's ability to capture past and future context increases the model's understanding of temporal dependencies and improves its predictive capabilities. For this reason, the proposed model has shown better performance.

**Table 5.** The Precision of different methods of NSL-KDD data.

| Attack type | **Proposed** | AIBPSF-IoMT | OMLIDS-PBIoT | AIMMFIDS |
| --- | --- | --- | --- | --- |
| DoS | **97** | 95 | 93 | 91 |
| R2L | **99** | 97 | 95 | 93 |
| Probe | **98** | 98 | 94 | 92 |
| U2R | **98** | 94 | 92 | 90 |
| Normal | **98** | 96 | 96 | 94 |
| Average | **98** | 96 | 94 | 92 |

**Table 6.** The Recall of different methods of NSL-KDD data.

| Attack type | **Proposed** | AIBPSF-IoMT | OMLIDS-PBIoT | AIMMFIDS |
| --- | --- | --- | --- | --- |
| DoS | **95** | 93 | 91 | 89 |
| R2L | **97** | 95 | 93 | 91 |
| Probe | **96** | 94 | 92 | 90 |
| U2R | **96** | 92 | 90 | 88 |
| Normal | **96** | 96 | 94 | 92 |
| Average | **96** | 94 | 92 | 90 |

In Figures 6, Table 7, and Table 8, a comparison is made between the proposed method and other methods in terms of precision, recall, accuracy, and F1-score on the UNSW-NB15 dataset. The proposed approach consistently outperforms alternative methods in evaluating intrusion detection methods on the UNSW-NB15 dataset. This is evident through examining precision, recall, accuracy, and F1-score metrics across various attack types. The proposed method exhibits remarkable precision, indicating high correctness in predicting attacks and elevated recall values, showcasing its proficiency in capturing a significant proportion of actual attacks. Comparatively, other methods, such as AIBPSF-IoMT, OMLIDS-PBIoT, and AIMMFIDS, though competitive, demonstrate slightly lower precision, recall, and accuracy values. The observed superiority of the proposed method can be attributed to several factors. The LSTM network's capacity to capture temporal dependencies in sequential data proves advantageous in the context of intrusion detection. Upon comparison with the NLS-KDD dataset, it becomes apparent that the precision, recall, and accuracy values in the UNSW-NB15 dataset are generally smaller. This discrepancy may be attributed to the increased complexity of the UNSW-NB15 dataset, characterized by diverse attack patterns and potential class imbalances. The proposed method, adept at handling these intricacies, demonstrates consistent and superior performance across various attack scenarios.

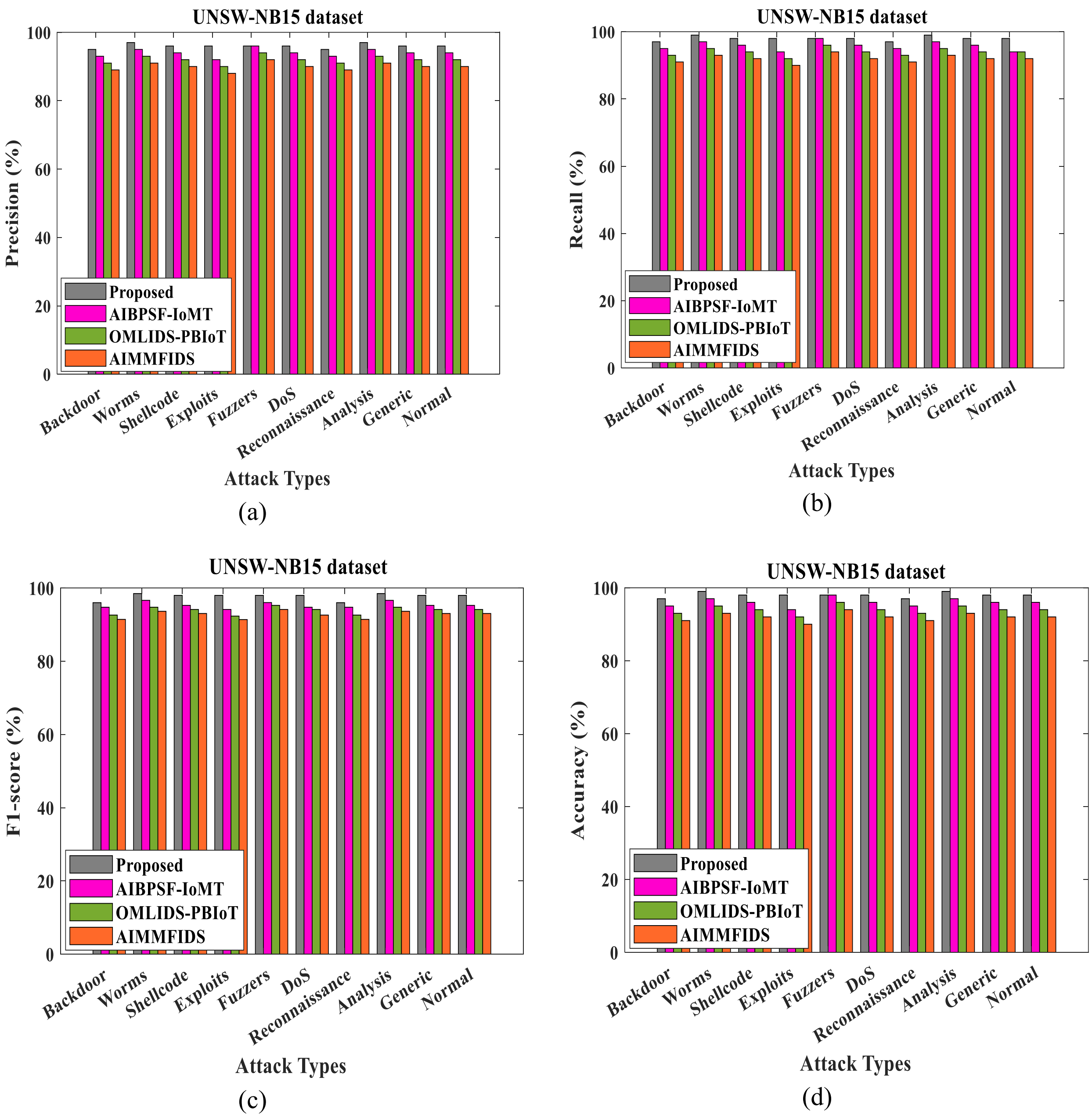

**Fig. 6** Comparative evaluation of the proposed method and three other methods in terms of precision, recall, F1-score, and accuracy on the UNSW-NB15 dataset.

**Table 7.** The Precision of different methods of UNSW-NB15 data.

| Attack type | **Proposed** | AIBPSF-IoMT | OMLIDS-PBIoT | AIMMFIDS |
|---|---|---|---|---|
| Backdoor | **95** | 94 | 92 | 90 |
| Worms | **97** | 96 | 94 | 92 |
| Shellcode | **96** | 95 | 93 | 93 |
| Exploits | **96** | 93 | 91 | 89 |
| Fuzzers | **96** | 97 | 95 | 91 |
| DoS | **96** | 95 | 93 | 91 |

| Reconnaissance | **95** | 94 | 92 | 90 |
|---|---|---|---|---|
| Analysis | **98** | 96 | 94 | 92 |
| Generic | **97** | 95 | 93 | 91 |
| Normal | **97** | 95 | 93 | 91 |
| Average | **97** | 95 | 93 | 91 |

**Table 8.** The Recall of different methods of UNSW-NB15 data.

| Attack type | **Proposed** | AIBPSF-IoMT | OMLIDS-PBIoT | AIMMFIDS |
|---|---|---|---|---|
| Backdoor | **94** | 92 | 90 | 88 |
| Worms | **96** | 94 | 92 | 90 |
| Shellcode | **95** | 93 | 91 | 89 |
| Exploits | **94** | 91 | 89 | 87 |
| Fuzzers | **95** | 95 | 93 | 91 |
| DoS | **95** | 93 | 91 | 89 |
| Reconnaissance | **96** | 92 | 90 | 88 |
| Analysis | **95** | 94 | 92 | 90 |
| Generic | **95** | 93 | 91 | 89 |
| Normal | **94** | 93 | 91 | 89 |
| Average | **96** | 93 | 91 | 89 |

Figure 7, Table 9, and Table 10 show the performance of different intrusion detection methods in terms of detection rate and false alarm rate in different attack percentages.

**Table 9.** The Detection rate of different methods of NSL-KDD data.

| Attack type | AP=30 | AP=40 | AP=50 | AP=60 | AP=70 | AP=80 |
|---|---|---|---|---|---|---|
| **Proposed** | **0.98** | **0.97** | **0.95** | **0.92** | **0.89** | **0.87** |
| AIBPSF-IoMT | 0.96 | 0.94 | 0.92 | 0.90 | 0.85 | 0.82 |
| OMLIDS-PBIoT | 0.96 | 0.93 | 0.90 | 0.86 | 0.83 | 0.81 |
| AIMMFIDS | 0.94 | 0.93 | 0.89 | 0.85 | 0.82 | 0.79 |

**Table 10.** The Detection rate of different methods of UNSW-NB15 data.

| Attack type | AP=30 | AP=40 | AP=50 | AP=60 | AP=70 | AP=80 |
|---|---|---|---|---|---|---|
| **Proposed** | **0.98** | **0.95** | **0.93** | **0.91** | **0.88** | **0.85** |
| AIBPSF-IoMT | 0.95 | 0.93 | 0.90 | 088 | 0.84 | 0.81 |
| OMLIDS-PBIoT | 0.95 | 0.92 | 0.89 | 0.85 | 0.83 | 0.81 |
| AIMMFIDS | 0.93 | 0.93 | 0.88 | 0.85 | 0.80 | 0.78 |

The proposed method consistently demonstrates a high detection rate, showcasing its effectiveness in identifying intrusions even as the percentage of attacks increases. Despite a rise in the false alarm rate at higher attack percentages, the method maintains a commendable balance between accurate detections and false positives. AIBPSF-IoMT exhibits a reasonable detection rate, though slightly lower than the proposed method, and a moderate false alarm rate. Similar to the proposed method, the detection rate decreases with an increase in attack percentages, and the false alarm rate sees a noticeable uptick at higher attack percentages, indicating some instances of false positives. OMLIDS-PBIoT maintains a moderate to high detection rate, showing resilience to varying attack percentages. However, like AIBPSF-IoMT, the false alarm rate increases gradually as the attack percentage rises. AIMMFIDS, while effective in detection, experiences a more pronounced decrease in the detection rate as the attack percentage increases. Furthermore, the false alarm rate for AIMMFIDS rises significantly at higher attack percentages, indicating a higher likelihood of false positives.

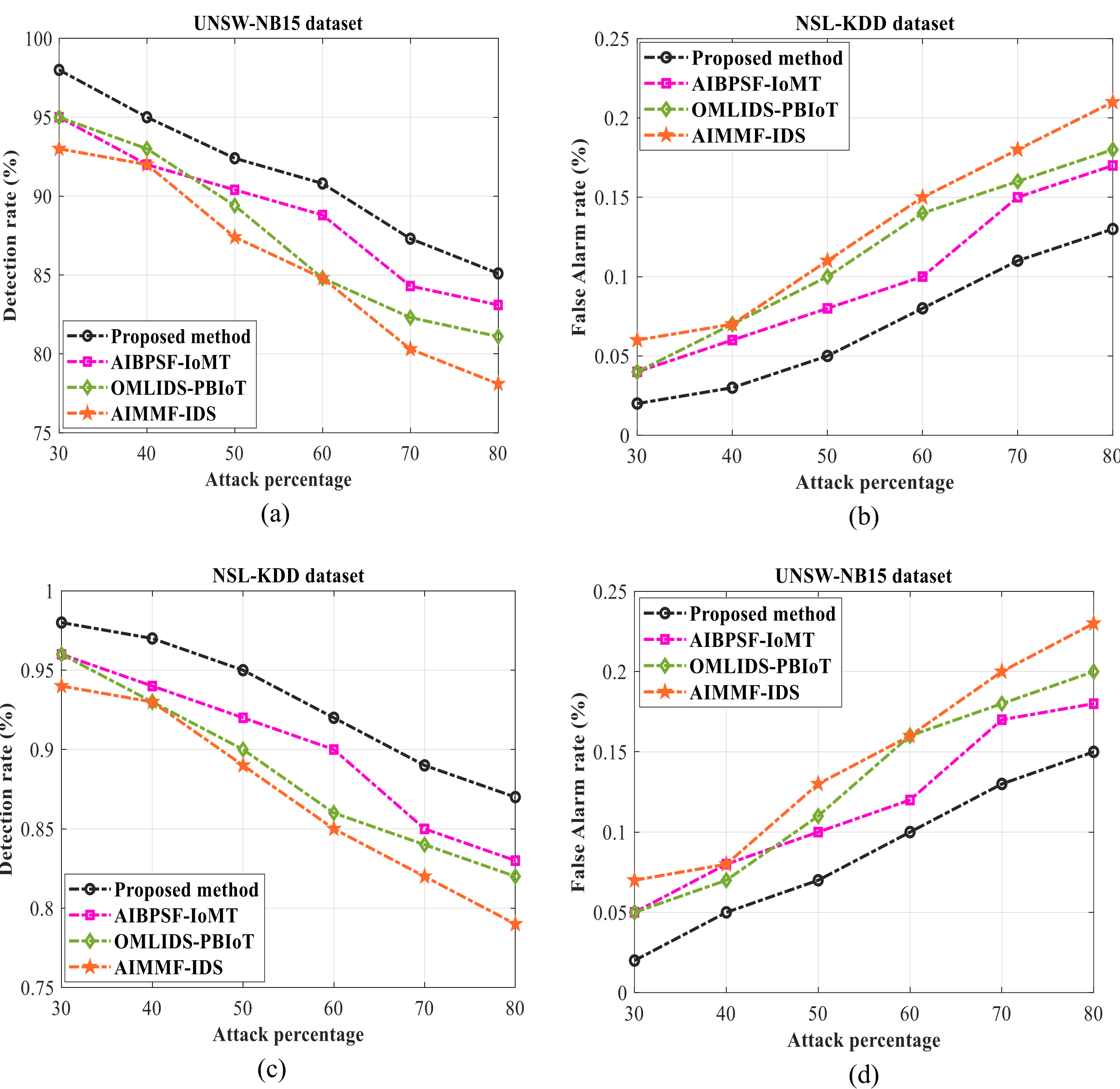

**Fig. 7** Detection rate and false alarm rate at varying attack levels for NSL-KDD and UNSW-NB15.

Figure 8 presents the confusion matrices for the BoT-IoT and DS2OS datasets. As evident from the visualizations, the proposed method demonstrates exceptional performance on both datasets, further highlighting its effectiveness and reliability.

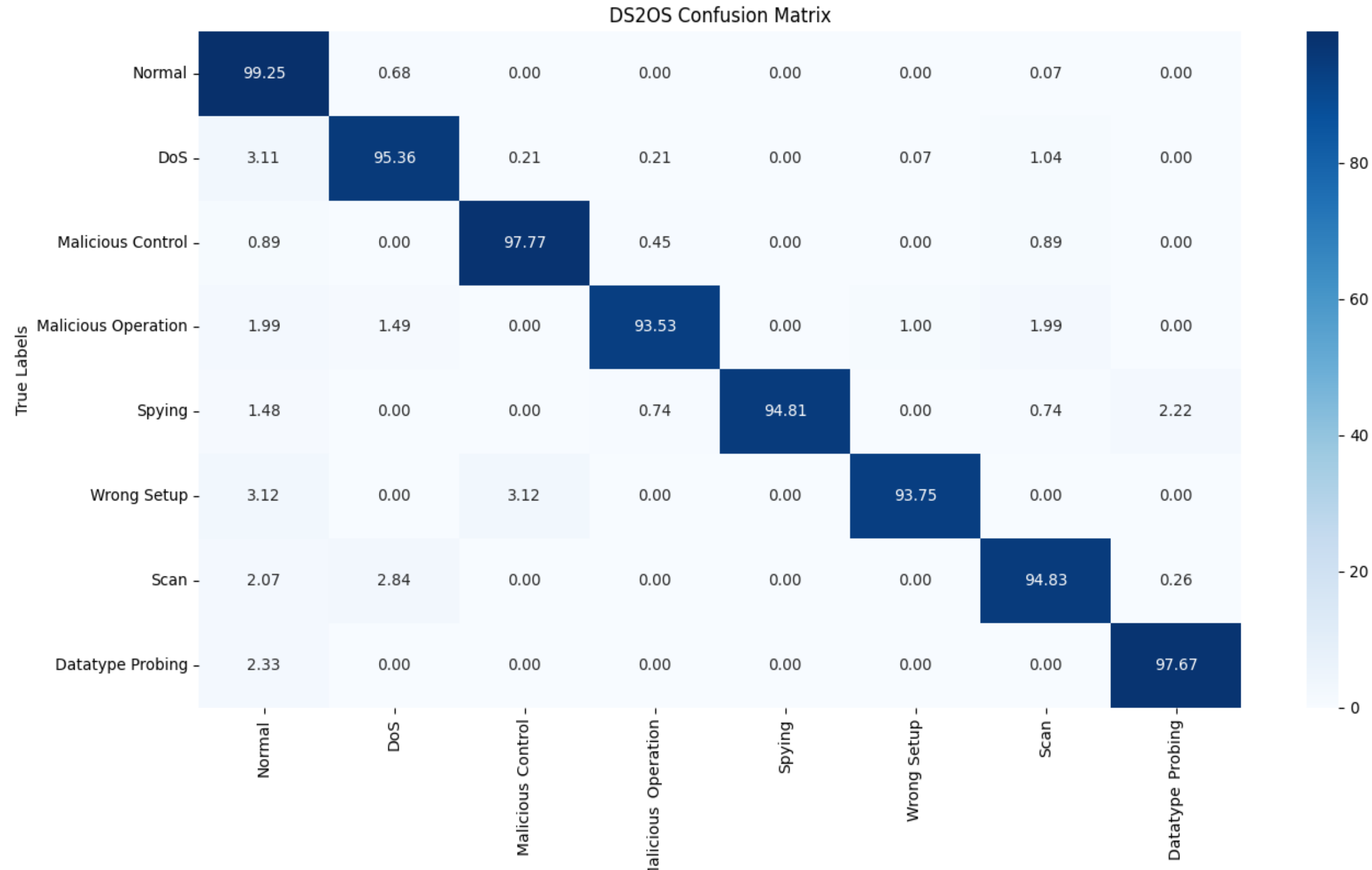

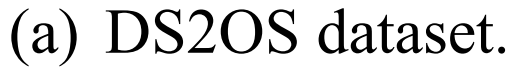

(a) DS2OS dataset.

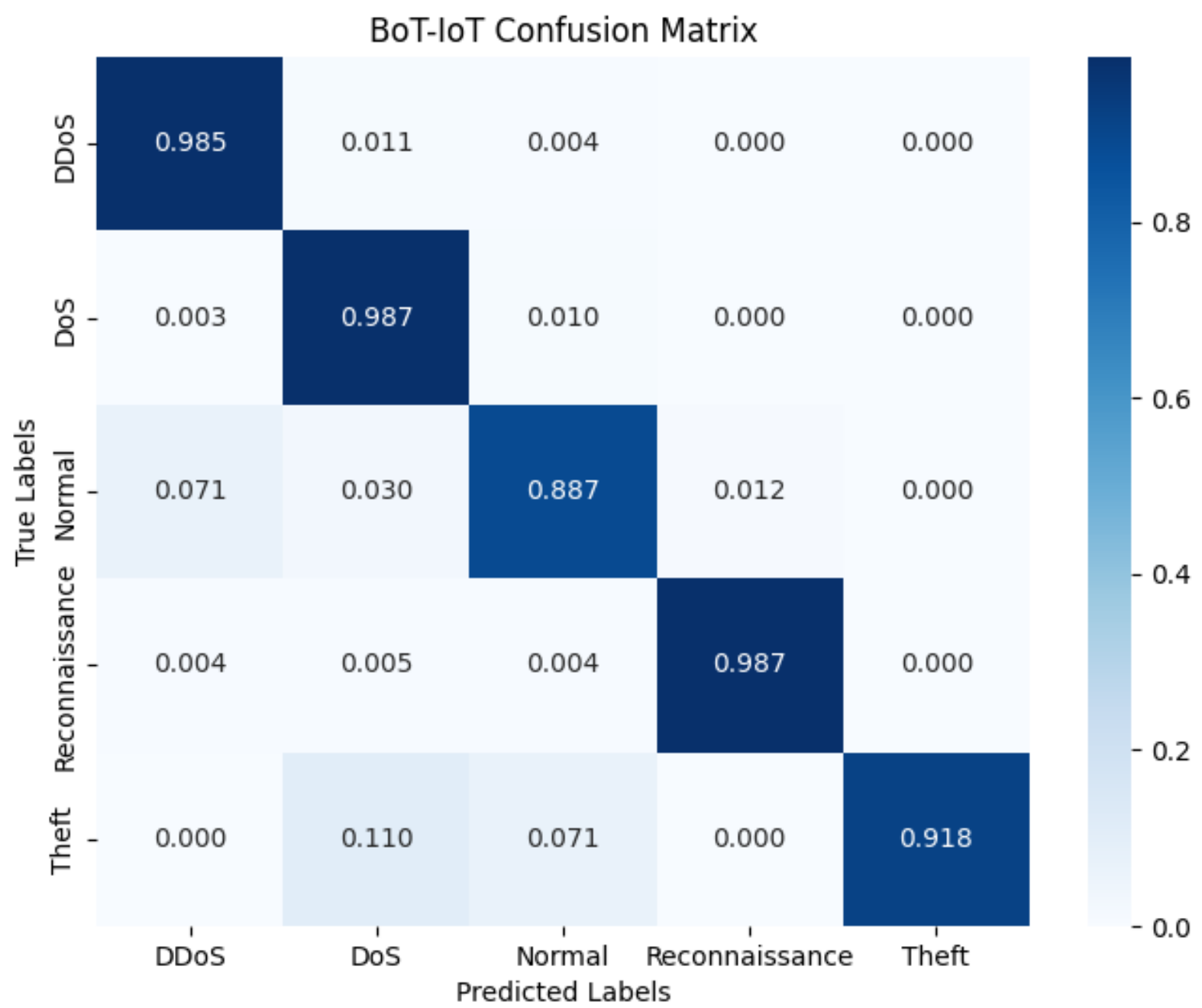

(b) BoT-IoT dataset.

**Fig. 8.** Confusion Matrix of (a) DS2OS and (b) BoT-IoT datasets.

## 4.3 Discussion

In this section, we provide a detailed discussion on the scalability of the proposed framework, the challenges associated with its real-world implementation, and the artificial intelligence algorithms employed.

**Scalability:**

- The proposed framework incorporates a lightweight proof-of-work (PoW) mechanism that reduces computational requirements, making it suitable for IoT environments with limited resources.
- By employing off-chain data storage using the InterPlanetary File System (IPFS), we minimize the on-chain data size, ensuring that the blockchain remains scalable even with a growing number of transactions and connected devices.

**Interoperability:**

- The modular design of the framework facilitates its integration with various existing systems. Each phase (trust management, blockchain-based privacy, and anomaly detection) can function independently or as part of the entire framework, allowing flexible deployment.
- The use of widely supported protocols and APIs enables seamless communication with existing IoT infrastructure and healthcare systems, ensuring adaptability to diverse environments.

**Computational Overhead:**

- We mitigate the computational burden by optimizing the blockchain and AI components. For instance, the lightweight LSTM model is designed to be computationally efficient while maintaining high accuracy in anomaly detection.
- Singular Value Decomposition (SVD) is applied to reduce the computational complexity of matrix operations in the LSTM, further enhancing efficiency.
- The decentralized fog computing architecture distributes the workload, reducing the strain on individual nodes and ensuring smooth operation even in resource-constrained scenarios.

**AI Models and Algorithms for Cyberattack Detection**

The proposed framework employs a lightweight Long Short-Term Memory (LSTM) network for anomaly detection, specifically designed to handle the constraints of IoT-based healthcare systems. The LSTM model effectively captures temporal dependencies in sequential data, enabling accurate detection of sophisticated cyberattacks, including previously unseen patterns. Additionally, Variational Autoencoders (VAEs) are used for data transformation, enhancing data privacy by generating secure representations of sensitive information.

**Limitations and Trade-offs**: While the LSTM and VAE models provide high accuracy and robustness, they come with certain limitations:

- **Computational Complexity**: Despite optimization through Singular Value Decomposition (SVD), the LSTM model may still present challenges in real-time applications with limited computational resources.
- **False Positives**: The trade-off between detection accuracy and false alarm rates may occasionally lead to higher false positives in more complex datasets.
- **Model Interpretability**: As with many AI-driven systems, the black-box nature of the models can make it difficult to interpret their decisions fully.

**Limitations of Blockchain Technology**

The integration of blockchain technology offers unparalleled data integrity and security; however, it introduces some challenges:

- **Latency**: The consensus mechanism, even with a lightweight Proof-of-Work (PoW), can lead to increased latency, particularly in time-sensitive healthcare applications.
- **Resource Requirements**: Blockchain nodes require sufficient computational power and storage capacity, which may strain IoT devices with limited resources.
- **Scalability**: As the number of devices and transactions grows, the blockchain network may experience scalability issues.

To address these limitations, the proposed framework incorporates off-chain storage via the InterPlanetary File System (IPFS) and fog computing architecture to reduce on-chain data volume and distribute computational workload. These enhancements ensure that blockchain integration remains feasible in IoT environments.

**Challenges in Deployment and Strategies to Address Them**

Deploying the proposed framework in real-world settings poses several challenges:

- **Heterogeneity of IoT Devices**: IoT environments often consist of devices with varying computational capabilities and protocols. To address this, the framework is designed with modular components and uses widely accepted standards and APIs for interoperability.
- **Network Latency**: Real-time applications require low-latency solutions, and fog computing nodes placed near IoT devices mitigate this issue by processing data locally before communicating with the cloud.
- **Cybersecurity Threat Evolution**: The dynamic nature of cyberattacks requires continuous updates to the AI models and blockchain configurations to adapt to emerging threats. Regular retraining of the models and employing explainable AI techniques will be integral to maintaining system effectiveness.
- **Cost and Resource Constraints**: Efficient resource allocation and lightweight algorithms ensure that the framework remains viable for deployment even in resource-constrained environments.

**Real-Time Application Challenges**: Real-time applications in healthcare demand low-latency and high-availability systems. The primary challenges include:

- **Latency**: Blockchain consensus mechanisms may delay data validation. This is addressed by optimizing the lightweight PoW and prioritizing critical transactions.
- **Scalability**: The modular fog and cloud architecture distributes processing loads, reducing bottlenecks and improving scalability.
- **Adaptability**: The system must adapt to rapidly evolving cyber threats without disrupting ongoing operations. AI models are periodically retrained with up-to-date datasets to address this challenge.

By addressing these challenges and aligning the framework with compliance standards, the proposed solution is positioned as a robust and scalable system for securing IoT-based healthcare environments in real-world scenarios.

## 5 Conclusion and Future Work

This paper introduces an innovative approach to address the critical challenge of medical data security in the IoT landscape. Our method seamlessly integrates BC technology and advanced machine learning techniques, providing a robust framework for enhancing the confidentiality and integrity of sensitive healthcare information. We proposed a comprehensive approach with three distinct phases to address the urgent need to improve security in IoT-based healthcare systems. In the first phase, we evaluated the reliability of IoT devices through reputation-based trust estimation and off-chain data storage. The second phase uses blockchain technology to prevent attacks and authenticate data. In the third phase, artificial intelligence is used to classify attacks. We compared the performance of the proposed method with three methods: AIBPSF-IoMT, OMLIDS-PBIoT, and AIMMFIDS. We achieved significant improvement in various metrics: precision (2%), accuracy (2%), recall (2%), attack detection rate (5%) and false alarm rate (3%). In future studies, we plan to address several critical areas to enhance the effectiveness and applicability of the proposed framework. First, the incorporation of explainable artificial intelligence (XAI) techniques will be explored to improve the system's robustness against adversarial attacks. By providing insights into the decision-making processes of the AI models, this approach aims to detect and mitigate subtle adversarial inputs that traditional methods might overlook. Second, we intend to expand the diversity of datasets used in experiments by including real-world healthcare IoT data. This will validate the framework's generalizability across various environments and ensure its applicability to a wide range of healthcare scenarios and attack patterns. Third, scalability will be further enhanced through the development of dynamic resource allocation strategies. These strategies will optimize processing power and bandwidth across the fog and cloud layers, and federated learning techniques will be investigated to decentralize model training and minimize data transfer overhead.

## Declarations

**Funding**

No funds

**Conflicts of interest**

Conflict of Interest is not applicable in this work.

**Availability of data and material**

Not applicable

**Ethical approval**

Not applicable

**CRediT author statement**

**Behnam Rezaei Bezanjani:** Conceptualization, Data curation, Investigation, Methodology, Resources, Software, Validation, Visualization, Writing - original draft, Writing - review & editing.

**Seyyed Hamid Ghafouri and Reza Gholamrezaei:** Formal analysis, Investigation, Methodology, Resources, Software, Supervision, Visualization, Writing - review & editing.

# Reference

1. Saheed, Y. K., & Misra, S. (2024). A voting gray wolf optimizer-based ensemble learning model for intrusion detection in the IoT. *International Journal of Information Security*, 1-25.
2. Saheed, K.Y., Usman, A.A., Sukat, F.D., & Abdulrahman, M. (2023). A novel hybrid autoencoder and modified particle swarm optimization feature selection for intrusion detection in the IoT network. *Frontiers of Computer Science*.
3. Khubrani, M. M. (2023). Artificial Rabbits Optimizer with Deep Learning Model for Blockchain-Assisted Secure Smart Healthcare System. *International Journal of Advanced Computer Science and Applications*, *14*(9).
4. Cai, J., Liang, W., Li, X., Li, K., Gui, Z., & Khan, M. K. (2023). GTxChain: a secure IoT smart blockchain architecture based on graph neural network. *IEEE IoT Journal*.
5. Kumar, P., Kumar, R., Srivastava, G., Gupta, G. P., Tripathi, R., Gadekallu, T. R., & Xiong, N. N. (2021). PPSF: A privacy-preserving and secure framework using blockchain-based machine-learning for IoT-driven smart cities. *IEEE Transactions on Network Science and Engineering*, *8*(3), 2326-2341.
6. Shrestha, A., & Mahmood, A. (2019). Review of deep learning algorithms and architectures. *IEEE access*, *7*, 53040-53065.
7. Alzubaidi, L., Zhang, J., Humaidi, A. J., Al-Dujaili, A., Duan, Y., Al-Shamma, O., ... & Farhan, L. (2021). Review of deep learning: Concepts, CNN architectures, challenges, applications, future directions. *Journal of big Data*, *8*, 1-74.

8. Ayyoubzadeh, S. M., Ayyoubzadeh, S. M., Zahedi, H., Ahmadi, M., & Kalhori, S. R. N. (2020). Predicting COVID-19 incidence through analysis of google trends data in Iran: data mining and deep learning pilot study. *JMIR public health and surveillance*, *6*(2), e18828.
9. Ahmad, R. W., Salah, K., Jayaraman, R., Yaqoob, I., Ellahham, S., & Omar, M. (2021). The role of blockchain technology in telehealth and telemedicine. *International journal of medical informatics*, *148*, 104399.
10. Alshammari, B. M. (2023). AIBPSF-IoMT: Artificial Intelligence and Blockchain-Based Predictive Security Framework for IoMT Technologies. *Electronics*, *12*(23), 4806.
11. Afaq, Y., & Manocha, A. (2023). Blockchain and Deep Learning Integration for Various Application: A Review. *Journal of Computer Information Systems*, 1-14.
12. Shafay, M., Ahmad, R. W., Salah, K., Yaqoob, I., Jayaraman, R., & Omar, M. (2023). Blockchain for deep learning: review and open challenges. *Cluster Computing*, *26*(1), 197-221.
13. Mudassir, M., Unal, D., Hammoudeh, M., & Azzedin, F. (2022). Detection of botnet attacks against industrial IoT systems by multilayer deep learning approaches. *Wireless Communications and Mobile Computing*, *2022*.
14. Azbeg, K., Ouchetto, O., & Andaloussi, S. J. (2022). BlockMedCare: A healthcare system based on IoT, Blockchain and IPFS for data management security. *Egyptian Informatics Journal*, *23*(2), 329-343.
15. Sudhakaran, P. (2022). Energy efficient distributed lightweight authentication and encryption technique for IoT security. *International Journal of Communication Systems*, *35*(2), e4198.
16. Yin, C., Zhang, S., Wang, J., & Xiong, N. N. (2020). Anomaly detection based on convolutional recurrent autoencoder for IoT time series. *IEEE Transactions on Systems, Man, and Cybernetics: Systems*, *52*(1), 112-122.
17. Xu, Z., Liang, W., Li, K. C., Xu, J., & Jin, H. (2021). A blockchain-based roadside unit-assisted authentication and key agreement protocol for internet of vehicles. *Journal of Parallel and Distributed Computing*, *149*, 29-39.
18. Liang, W., Ning, Z., Xie, S., Hu, Y., Lu, S., & Zhang, D. (2021). Secure fusion approach for the IoT in smart autonomous multi-robot systems. *Information Sciences*, *579*, 468-482.
19. Khalaf, O. I., & Abdulsahib, G. M. (2021). Optimized dynamic storage of data (ODSD) in IoT based on blockchain for wireless sensor networks. *Peer-to-Peer Networking and Applications*, *14*, 2858-2873.
20. Srinivasu, P. N., Bhoi, A. K., Nayak, S. R., Bhutta, M. R., & Woźniak, M. (2021). Blockchain technology for secured healthcare data communication among the non-terminal nodes in IoT architecture in 5G network. *Electronics*, *10*(12), 1437.
21. Fotohi, R., & Pakdel, H. (2021). A lightweight and scalable physical layer attack detection mechanism for the internet of things (IoT) using hybrid security schema. *Wireless personal communications*, *119*(4), 3089-3106.
22. Zhang, J. (2021). Distributed network security framework of energy internet based on IoT. *Sustainable Energy Technologies and Assessments*, *44*, 101051.
23. Fotohi, R., & Effatparvar, M. (2013). A cluster based job scheduling algorithm for grid computing. *International Journal of Information Technology and Computer Science (IJITCS)*, *5*(12), 70-77.
24. Fotohi, R., Aliee, F. S., & Farahani, B. (2024). Decentralized and robust privacy-preserving model using blockchain-enabled federated deep learning in intelligent enterprises. *Applied Soft Computing*, *161*, 111764.

25. Zheng, Z., Xie, S., Dai, H. N., Chen, W., Chen, X., Weng, J., & Imran, M. (2020). An overview on smart contracts: Challenges, advances and platforms. *Future Generation Computer Systems*, *105*, 475-491.
26. Ebazadeh, Y., & Fotohi, R. (2022). A reliable and secure method for network‐layer attack discovery and elimination in mobile ad‐hoc networks based on a probabilistic threshold. Security and Privacy, 5(1), e183.
27. Li, Y., Chen, C., Liu, N., Huang, H., Zheng, Z., & Yan, Q. (2020). A blockchain-based decentralized federated learning framework with committee consensus. *IEEE Network*, *35*(1), 234-241.
28. Fotohi, R., Aliee, F. S., & Farahani, B. (2024). A lightweight and secure deep learning model for privacy-preserving federated learning in intelligent enterprises. IEEE Internet of Things Journal, 11(19), 31988-31998.
29. Philip, A. O., & Saravanaguru, R. K. (2020). Secure incident & evidence management framework (SIEMF) for internet of vehicles using deep learning and blockchain. *Open Computer Science*, *10*(1), 408-421.
30. AlKadi, O., Moustafa, N., Turnbull, B., & Choo, K. K. R. (2019). Mixture localization-based outliers models for securing data migration in cloud centers. *IEEE Access*, *7*, 114607-114618.
31. Liang, G., Weller, S. R., Luo, F., Zhao, J., & Dong, Z. Y. (2018). Distributed blockchain-based data protection framework for modern power systems against cyber attacks. *IEEE Transactions on Smart Grid*, *10*(3), 3162-3173.
32. Liu, X., Muhammad, K., Lloret, J., Chen, Y. W., & Yuan, S. M. (2019). Elastic and cost-effective data carrier architecture for smart contract in blockchain. *Future Generation Computer Systems*, *100*, 590-599.
33. Rathore, S., & Park, J. H. (2018). Semi-supervised learning based distributed attack detection framework for IoT. *Applied Soft Computing*, *72*, 79-89.
34. Wang, Z. (2018). Deep learning-based intrusion detection with adversaries. *IEEE Access*, *6*, 38367-38384.
35. Mirjalili, S., & Lewis, A. (2016). The whale optimization algorithm. *Advances in engineering software*, *95*, 51-67.
36. Tavallaee, M., Bagheri, E., Lu, W., & Ghorbani, A. A. (2009, July). A detailed analysis of the KDD CUP 99 data set. In *2009 IEEE symposium on computational intelligence for security and defense applications* (pp. 1-6). Ieee.
37. Moustafa, N., & Slay, J. (2015, November). UNSW-NB15: a comprehensive data set for network intrusion detection systems (UNSW-NB15 network data set). In *2015 military communications and information systems conference (MilCIS)* (pp. 1-6). IEEE.
38. Al-Qarafi, A., Alrowais, F., S. Alotaibi, S., Nemri, N., Al-Wesabi, F. N., Al Duhayyim, M., ... & Al-Shabi, M. (2022). Optimal machine learning based privacy preserving blockchain assisted IoT with smart cities environment. *Applied Sciences*, *12*(12), 5893.
39. Alohali, M. A., Al-Wesabi, F. N., Hilal, A. M., Goel, S., Gupta, D., & Khanna, A. (2022). Artificial intelligence enabled intrusion detection systems for cognitive cyber-physical systems in industry 4.0 environment. *Cognitive Neurodynamics*, *16*(5), 1045-1057.
40. Bezanjani, B. R., Ghafouri, S. H., & Gholamrezaei, R. (2024). Fusion of machine learning and blockchain-based privacy-preserving approach for healthcare data in the Internet of Things. The Journal of Supercomputing, 80(17), 24975-25003.
41. Kumar, R., Khan, S. A., Alharbe, N., & Khan, R. A. (2024). Code of silence: cyber security strategies for combating deepfake disinformation. *Computer Fraud & Security*, *2024*(4).

42. Sahu, K., & Kumar, R. (2024). A secure decentralised finance framework. *Computer Fraud & Security*, *2024*(3).
43. Sahu, K., & Srivastava, R. K. (2021). Predicting software bugs of newly and large datasets through a unified neuro-fuzzy approach: Reliability perspective. *Advances in Mathematics: Scientific Journal*, *10*(1), 543-555.
44. Kumar, R., Khan, S. A., & Khan, R. A. (2023). *Software Durability: Concepts and Practices*. CRC Press.
45. Kumar, R., Baz, A., Alhakami, H., Alhakami, W., Baz, M., Agrawal, A., & Khan, R. A. (2020). A hybrid model of hesitant fuzzy decision-making analysis for estimating usable-security of software. *IEEE Access*, *8*, 72694-72712.
46. Kumar, R., & Mishra, S. K. (2024). Assessing the impact of heat vulnerability on urban public spaces using a fuzzy-based unified computational technique. *AI & SOCIETY*, 1-18.
47. Kumar, R., & Khan, R. A. (2024). Securing communication protocols in military computing. *Network Security*, *2024*(3).
48. Pandey, A. K., Das, A. K., Kumar, R., & Rodrigues, J. J. (2024). Secure Cyber Engineering for IoT-Enabled Smart Healthcare System. *IEEE Internet of Things Magazine*, *7*(2), 70-77.
49. Kumar, R., & Ahmad Khan, R. (2024). Securing military computing with the blockchain. *Computer Fraud & Security*, *2024*(2).
50. Ch, R. (2024, June). Remix-based Real Time Blood Bank Communication Integrating Access Control Using XGBoost and Supabase. In *2024 IEEE Students Conference on Engineering and Systems (SCES)* (pp. 1-6). IEEE.
51. Sree, G. D., Rupa, C., & Gayathri, U. M. L. (2024, July). A novel substantial cipher against passive attacks using fuzzy logic based approach. In *AIP Conference Proceedings* (Vol. 3075, No. 1). AIP Publishing.
52. Guggilam, N. V. R., Chiramdasu, R., Nambur, A. B., Mikkineni, N., Zhu, Y., & Gadekallu, T. R. (2024). An expert system for privacy-driven vessel detection harnessing YOLOv8 and strengthened by SHA-256. *Computers & Security*, *143*, 103902.
53. Rupa, C., Sultana, S. A., Malleswari, R. P., Dedeepya, C., Gadekallu, T. R., Song, H. K., & Piran, M. J. (2024). IOMT Privacy Preservation: A Hash-Based DCIWT approach for detecting tampering in medical data. *IEEE Access*.
54. Fotohi, R., Aliee, F. S., & Farahani, B. (2022, November). Federated learning: Solutions, challenges, and promises. In *2022 6th Iranian Conference on Advances in Enterprise Architecture (ICAEA)* (pp. 15-22). IEEE.
55. Rupa, C., MidhunChakkarvarthy, D., Patan, R., Prakash, A. B., & Pradeep, G. G. (2022). Knowledge engineering–based DApp using blockchain technology for protract medical certificates privacy. *IET Communications*, *16*(15), 1853-1864.